# Analysis of localized Schmidt decomposition modes and of entanglement in atomic and optical quantum systems with continuous variables

A.Yu. Bogdanov**, Yu.I. Bogdanov*, K.A. Valiev*

*Institute of Physics and Technology, Russian Academy of Sciences*
**Faculty of Physics, Moscow M.V. Lomonosov State University, Russia*

We investigate the procedure of Schmidt modes extraction in systems with continuous variables. An algorithm based on singular value matrix decomposition is applied to the study of entanglement in an "atom-photon" system with spontaneous radiation. Also, this algorithm is applied to the study of a bi-photon system with spontaneous parametric down conversion with type-II phase matching for broadband pump.

We demonstrate that dynamic properties of entangled states in an atom-photon system with spontaneous radiation are defined by a parameter equal to the product of the fine structure constant and the atom-electron mass ratio. We then consider the evolution of the system during radiation and show that the atomic and photonic degrees of freedom are entangling for the times of the same order of magnitude as the excited state life-time. Then the degrees of freedom are de-entangling and asymptotically approach to the level of small residual entanglement that is caused by momentum dispersion of the initial atomic packet.

Finally, we investigate the process of coherence loss between modes in type-II parametric down conversion that is caused by non-linear crystal properties.

## 1 Introduction

Schmidt decomposition is an effective tool for analyzing entangled quantum states [1]. Recently, the problem of entanglement in quantum systems with continuous variables has been widely discussed [2,3,4]. It appears that despite the infinite large number of degrees of freedom, in some interesting cases such systems can be effectively described by a relatively small number of Schmidt modes.

The main goal of our research is to develop numerical methods of describing entangled systems by the means of singular value decomposition (SVD). The algorithm developed was applied to describing entanglement in an atom-photon system with spontaneous atomic radiation and also to a similar research of a bi-photon system with spontaneous parametric down conversion with type-II phase matching.

For the first time it was shown that the evolution of entangled states in an atom-photon system is defined by a parameter, equal to the product of the fine structure constant and the atom-electron mass ratio. The dynamics of the system is defined by a superposition of the initial and the developing states. It is shown that the atomic and photonic degrees of freedom during radiation are "entangling" for the times of about the life-time of the excited state. Then the degrees of freedom are "de-entangling" and asymptotically approach to the small residue level of entanglement that is due to the energetic dispersion of the initial wave atomic packet (the square dispersion of the momentum). An analytical structure of photonic modes is presented. An interpretation of the derived results is given.

An analysis of spontaneous parametric down conversion with type-II matching is provided. The analysis is based on polarized density matrix and a coherence parameter. The process of coherence loss between modes is considered. That loss is due to the different properties of an ordinary and an extraordinary photons in a non-linear crystal. The dependence of the coherence between modes on the product of the crystal length and the pump bandwidth is given. Some examples of the Schmidt modes for various coherence levels are provided.

The structure of our paper is as follows. In section 2 we give some general information on the Schmidt decomposition modes and the numerical algorithm based on singular value matrix decomposition. Then we compare the developed approach with the methods used in other papers.

In section 3 we consider entangled states of the system atom-photon with spontaneous radiation. We provide a description of the structure of Schmidt modes, the Schmidt number, the entropy of entanglement for the different atomic packet impulse spread.

In section 4 asymptotical properties of the atom-photon system in momentum representation are studied. The study is based on the Weisskopf- Wigner theory of natural spectral line width of a two-level atom. Furthermore, the research concerns the recoil energy of the atom and momentum distribution of the initial wave packet.

In Section 5 we consider spontaneous parametric down conversion of light. That implies that the initial photon splits in two photons each carrying about one half of its energy. The entangled frequency-polarization properties of such systems are evident when one uses impulse (10-100 fs) laser pump in the process of SPDC with the type-II phase matching. In that process the emerging photons are perpendicularly polarized to one another.

In Section 6 we investigate the polarized density matrix for the case of SPDC with the second type phase matching. We introduce a coherence parameter for two spatially separated modes and analyze its dependence on the product of the crystal length and the pump bandwidth.

In section 7 we summarize the results of our present work.

## 2. Schmidt decomposition

Let the probability amplitude (wave function) of the system $\psi(p,q)$ be a function of two continuous variables $p$ and $q$. In numerical calculations the function can be represented in a discrete form as a matrix $\psi_{j_1 j_2} = \psi(p_{j_1}, q_{j_2})$, where $1 \leq j_1 \leq n$, $1 \leq j_2 \leq n$. We suppose that for the function the uniform discrete computational mesh of size $n \times n$ is defined. The number $n$ can be set quite large. For one it is sufficient to take n equal to about 30-50. Nevertheless we obtain our final results for $n$ equal to 300-800.

The matrix $\psi(p_{j_1}, q_{j_2})$ can be represented in Schmidt decomposition form [1]:

$$\psi(p_{j_1}, q_{j_2}) = \sum_{k=1}^{n} \sqrt{\lambda_k} \psi_k^{(1)}(p_{j_1}) \psi_k^{(2)}(q_{j_2}) \qquad (1)$$

where $\lambda_k$ are the weight factors that meet the normalization condition.

$$\sum_k \lambda_k = 1 \qquad (2)$$

We assume that the summands in decomposition (1) are represented in the order of non-increasing of the coefficients $\lambda_k$.

The functions $\psi_k^{(1)}(p)$ и $\psi_k^{(2)}(q)$ are called the Schmidt modes. Generally speaking, the number of modes is equal to the discretization number $n$. Even so, in many cases that are of practical interest one can limit oneself to considering only a few primary modes. That is due to the fact that its total weight is close to 1.

In mathematical terms the possibility to consider only a few primary modes implies that one dramatically decreases the problem dimension. Thus the infinite Hilbert space is effectively substituted with a space of relatively small dimension.

If in (1) only the first summand is considerably large, then the degrees of freedom that correspond to the variables $p$ and $q$ are non-entangled. Otherwise, one has to take into consideration more that one summand and then the degrees of freedom become entangled. For instance, the registration of the variable $p$ in state $\psi_k^{(1)}(p)$ will imply that the variable $q$ is to be found in state $\psi_k^{(2)}(q)$ (for the same $k$).

We shall introduce some characteristics that describe the degree of entanglement. Let $\lambda_k$ be the spectra of eigenvalues in Schmidt decomposition. Then the effective number of modes K can be estimated as [2]:

$$K = \frac{1}{\sum_k \lambda_k^2} \qquad (3)$$

Due to its definition, the number K is no lower than one. It is equal to one only if there is a single non-zero mode in the decomposition.

One may consider the well-known entropy of entanglement as another measure of multi-modes [1].

$$S = -\sum_k \lambda_k \log_2 \lambda_k \qquad (4)$$

The entropy of entanglement is always non-negative. It is equal to zero only if there is a single non-zero mode in Schmidt decomposition.

The Schmidt decomposition is much more compact than the decomposition of psi- function in 2-D Fourier transform on an arbitrary set of orthogonal functions. The Fourier coefficients produce a rectangular (e.g. square) matrix. For the Schmidt decomposition that matrix becomes diagonal. Then the basis functions are no longer arbitrary. They become quite well-defined as their appearance characterizes the considered physical problem.

Let us describe the algorithm of numerical extraction of Schmidt modes in more detail. Let $\psi$ be a matrix of size $n \times n$ with elements $\psi_{j_1 j_2}$. Let us introduce matrix $M$ as follows:

$$M = \psi \cdot \psi^+ \tag{5}$$

If one calculates the eigenvalues and eigenfunctions of matrix $M$ then it will be represented as:

$$M = UDU^+, \tag{6}$$

Here $U$ is a unitary matrix that consists of eigenfunctions of matrix $M$. (Every column of matrix $U$ is an eigenfunction of $M$). Matrix $D$ is a diagonal matrix that consists of eigenvalues $\lambda_k$ of matrix $M$. Also, we consider that $\lambda_k$ are placed on the diagonal in non-increasing order.

The diagonal elements of matrix $D$ are the desired weight multipliers $\lambda_k$ of Schmidt decomposition. The mode $\psi_k^{(1)}(p)$ is given by the k-th **column** of matrix $U$.

In order to obtain modes $\psi_k^{(2)}(q)$ let us introduce matrix $V$ as:

$$V = \sqrt{D^{-1}} U^+ \psi \tag{7}$$

As a rule, in high-dimension problems matrix $D$ consists of elements that are very close to zero. This could lead to dividing by zero while calculating $D^{-1}$. To avoid such effect one may add small summands to the diagonal $D$ (e.g. of the order of magnitude $10^{-12}$ - $10^{-16}$). The results do not really depend on the level of smallness of the summands. They are only introduced to avoid the division by zero. One may also "cut" the matrix $D$ dimension to leave only $r$ non-zero diagonal elements $\lambda_1, \lambda_2, \ldots, \lambda_r$ (one also needs to leave only $r$ columns in the matrix $U$).

As a result to obtain mode $\psi_k^{(2)}(q)$ one needs only to take the $k$-th **row** of matrix $V$. Using the matrices $U$ and $V$, the matrix of the probability amplitudes $\psi$ can be written as:

$$\psi = U \cdot S \cdot V \tag{8}$$

where $S = \sqrt{D}$ is a diagonal matrix that has non-zero diagonal elements placed in non-increasing order. The decomposition (8) is a singular value decomposition, while the coefficients $\sqrt{\lambda_k}$ are the singular matrix values.

The algorithm shows that the calculation of Schmidt modes is a self-consistent by the variables $p$ and $q$ procedure. Every column of matrix $U$ (every mode $\psi_k^{(1)}(p)$) can take an unimportant phase multiplier. Nonetheless that multiplier makes the entangled mode phase change consistently.

The developed algorithm differs from other similar algorithms discussed in papers [3,4]. In paper [3] the Schmidt modes are derived from two independent equations ((3) and (4) in [3]). Such analysis ignores the aforementioned self-consistency of the problem and can be erroneous. In particular, in [3] the third and the fourth modes have a wrong relative phase $\pi$ for $\psi_k^{(1)}(p)$ and $\psi_k^{(2)}(q)$ (they have a wrong relative sign).

In paper [4] preliminary expansion of a two-variable function (probability amplitude) in a Fourier series based on Chebyshev- Hermite functions is assumed. Such consideration leads to unnecessary complication of the numerical calculation that causes loss of precision. (That is due to the fact that the authors in [4] limit themselves to considering not more than 20 functions for each variable).

### 3. The entangled states of atom-photon system with spontaneous radiation

We assume that the primary state of a trapped atom is described by a Gauss wave packet. That corresponds to the main state of harmonic oscillator. At the start time $t = 0$ the atom is excited by a short laser pulse and simultaneously the trap field is turned off, so that the atom freely evolutes. We consider that a two-level atom and we assume that the laser pulse duration is small compared to the excited state life-time $1/\gamma$ (where $\gamma$ stands for the excited level width).

Rigorous analysis of the atom-photon system states is based on the quantum theory of radiation and it is given in [5]. We shall limit ourselves to simple heuristic considerations that provide the right answer.

Let us consider the case when an atom and a photon move in opposite directions in one dimensional space. Then, by omitting unimportant multipliers, one gets the following approximation for the wave function of atom-photon system.

$$\psi(p,q) = \theta(\tau - p)\exp\left(-\frac{1}{2}(\tau - p)\right)\exp\left(-\frac{\eta^2(p+q)^2}{2(1+\tau\eta^2\xi_0 \cdot i)}\right) \quad (9)$$

where $\tau = \gamma t$ is the dimensionless time. The equation (9) is true if $\tau \gg 1$.

$$p = \frac{r_{ph}\gamma}{c}$$ - is the dimensionless coordinate of the photon (the coordinate of the photo detector)

$$q = \frac{r_{at}\gamma}{v_{rec}}$$ - is the dimensionless coordinate of the atom,.

where $v_{rec} = \frac{\hbar\omega}{Mc}$ is the recoil velocity of the atom.

We assume that both $r_{ph}$ and $r_{at}$ ($p$ and $q$) can be either positive or negative values. In Eq.(9) we ignore all the multipliers that do not depend on the coordinates of the atom and the photon.

$$\eta = \frac{v_{rec}}{\gamma a_0} = \frac{\hbar\omega}{Mc\gamma a_0}$$ - is a dimensionless value that is defined by the Gauss wave packet width $a_0$ at the start time.

The value $\eta$ is inversely proportional to the width of Gauss packet $a_0$. Therefore, due to the principle of uncertainty, $\eta$ is directly proportional to the momentum spread in the initial state. As a result, one may call $\eta$ the momentum spread parameter. A similar parameter was introduced in [2] where it was called the control parameter of the degree of entanglement.

It is assumed that at the first moment of time the state of the atom is a one-dimensional Gauss packet of the width $a_0$:

$$\psi(r_{at}, t=0) = const \cdot \exp\left(-\frac{r_{at}^2}{2a_0^2}\right) \quad (10)$$

We shall note that according to (10) the variance of coordinate distribution is $a_0^2/2$.

$$\xi_0 = \frac{Mc^2}{\hbar\omega}\frac{\gamma}{\omega}$$ - is a dimensionless value that consists only of the parameters that characterize the atomic system. This value is constant.

The state (9) of the atom-photon system has a simple physical interpretation. The first multiplier is the Heaviside step function that shows that the psi-function is equal to zero for $p > \tau$. The second multiplier describes an exponential decrease of the probability amplitude of the excited atom [6]. The third multiplier is the source of entanglement in the atom-photon system. Let us explain it in more detail. From the excitation time $\tau = 0$ till the registration time $\tau$ the atom and the radiated photon evolve as a whole. The evolution of the center of inertia of the system corresponds to the free movement of a Gauss wave packet that spreads in the course of time. In other words, according to the law of conservation of momentum, no matter what happens inside the system, the center of inertia evolves as a freely moving particle as long as there is no external excitation on the system (atom or photon registration).

In dimensionless coordinates the coordinate of the center of inertia is $p + q$. That is due to the fact that if $r_{ci}$ is the center of inertia coordinate, then the coordinates of the photon and the atom are (in 1D-approximation) $r_{ph} = r_{ci} + ct$, $r_{at} = r_{ci} - v_{rec}t$ as they move in opposite directions. Then for the dimensionless coordinate of the inertia center one gets:

$$p + q = \frac{r_{ph}\gamma}{c} + \frac{r_{at}\gamma}{v_{rec}} = r_{ci}\gamma\left(\frac{1}{c} + \frac{1}{v_{rec}}\right) \approx \frac{r_{ci}\gamma}{v_{rec}} \quad (11)$$

Also, it can be shown that the evolution of a freely spreading Gauss packet leads to the equation expressed by the third multiplier in (9) (e.g. see [7]).

The impossibility to fix the coordinates of the centre of inertia is of principal importance. If initially one takes a narrow packet (that corresponds to the small $a_0$) then it spreads rapidly due to the uncertainty relation. Then a small uncertainty of the coordinate of the centre of inertia leads to a large uncertainty of the momentum and, consequently, that leads to a large uncertainty of the coordinate later.

One may show that there is an optimal value of the initial packet width $a_0$ that corresponds to the minimum packet variance at the registration time $\tau$. It corresponds to the optimal value of $\eta$:

$$\eta_{opt} = \frac{1}{\sqrt{\xi_0 \tau}} \quad (12)$$

The evolution of the entangled states of atom-photon system is defined by above introduced parameter $\xi_0 = \frac{Mc^2}{\hbar\omega}\frac{\gamma}{\omega}$ that consists of atomic constants. The parameter $\xi_0$ is defined as the product of a large value $\frac{Mc^2}{\hbar\omega}$ (ratio of the internal energy $Mc^2$ to the photon energy $\hbar\omega$) and a small value $\frac{\gamma}{\omega}$ (ratio of the excited level width to the transition frequency). On the basis of formulas of quantum electrodynamics [6] in dipole approximation we get the following estimate for $\xi_0$:

$$\xi_0 = \frac{Mc^2}{\hbar\omega}\frac{\gamma}{\omega} \sim \frac{\alpha M}{m} = \frac{M}{137 m} \quad (13)$$

where $m$ - is the mass of the electron, $M$ - the mass of the atom, $\alpha = \frac{1}{137}$ - the fine structure constant.

It is evident that in real atomic systems $\xi_0$ is a high value ($\xi_0 > 10$) due to the fact that even the lightest atom, hydrogen, is about 2000 times heavier than the electron.

During the process of radiation the wave packet is spreading. Yet this spreading is small, which means that during the radiation time $t \sim \frac{1}{\gamma}$ the "diffusion" quantum length is small compared to the initial atomic packet length $a_0$ [5]. From that condition one easily gets:

$$\eta << \frac{1}{\sqrt{\xi_0}} \qquad (14)$$

Due to (13) the condition (14) implies that the parameter $\eta$ a priori is a small value (for the order of value 0.1 or smaller). Note that in papers [2, 8] large parameters $\eta \sim (10-1000)$ were considered. For the aforementioned reasons such consideration does not seem quite physically correct.

From the other point of view, the parameter $\eta$ can not be too small either. Small values of $\eta$ stand for large values of the atomic packet size. Consequently, that leads to the coherence loss between the radiating photon and the atomic oscillator. For large radiating atomic systems the phase multiplier $\langle \exp(i\vec{k}\vec{r}) \rangle$ is close to zero. The coherence condition between the radiated photon and the atomic oscillator imply that the size $a_0$ of the atomic packet is small compared to the wave length $\lambdabar = \lambda/(2\pi)$. From that one easily gets a condition:

$$\eta >> \frac{1}{\xi_0} \qquad (15)$$

Let us take into consideration that $\frac{a_0}{\lambdabar} = \frac{1}{\xi_0 \eta}$ due to the definition. Then the conditions (14) and (15) can be rewritten as:

$$\frac{1}{\sqrt{\xi_0}} << \frac{a_0}{\lambdabar} << 1 \qquad (16)$$

Thus, the conditions of applying the described theory imply the assignment of the initial atomic packet width in a narrow range, defined by (16).

On Fig. 1a an example of probability amplitudes for the compound atom-photon system is given (the first Schmidt mode). It is evident that the spatial atomic wave function localization is negligible (the large atomic packet width does not even allow one to predict the recoil direction of the atom). The figure corresponds to the following parameter values: $\xi_0 = 100$, $\eta = 0.03$, $\tau = 10$.

Fig 1b shows the structure of the primary three wave functions of the radiation. There the greatest part stands for the first Schmidt mode.

One may approximate the structure of the Schmidt modes by a set of orthogonal functions that is based on the Laguerre polynomials ($C_k$ is the normalization constant).

$$\psi_k^{(1)}(p) = C_k L_k(\tau - p) \exp\left(-\frac{1}{2}(\tau - p)\right) \theta(\tau - p) \quad (k = 0,1,...) \qquad (17)$$

The numerical calculation results of the Schmidt modes are very close to the results of the analytical approximation. The relative approximation precision is about $10^{-3}$. In our calculations we used a discrete model with the number of points $n = 800$.

According to the Weisskopf- Wigner model, the dynamics of the atom-photon system state during radiation is defined by the superposition of the initial and the developing states. The initial state weight $|e\rangle|0\rangle$ (the excited atom and zero photons) decreases with time as $\lambda^{(e)}(t) = \exp(-\gamma t)$. At the same time, the weight of the developing state $|g\rangle|1\rangle$ (the atom in the ground state plus one photon) increases as $\lambda^{(g)}(t) = 1 - \exp(-\gamma t)$. For that reasons, the Schmidt number and the entanglement entropy in zero approximation (without taking account of the fine structure of the state (9)) are defined as:

$$K_0(t) = \frac{1}{\left(\lambda^{(e)}(t)\right)^2 + \left(\lambda^{(g)}(t)\right)^2} \tag{18}$$

$$S_0(t) = -\left(\lambda^{(e)}(t)\right)^2 \log_2\left\{\left(\lambda^{(e)}(t)\right)^2\right\} - \left(\lambda^{(g)}(t)\right)^2 \log_2\left\{\left(\lambda^{(g)}(t)\right)^2\right\} \tag{19}$$

Numerical calculations (Fig. 2a) shoe that considering the fine structure does not have almost impact on the estimates, given by the Eq. (18) and (19).

The Figures shoe that the atomic and the photonic degrees of freedom during radiation are entangling for the time $\tau < 1$ and are de-entangling for $\tau \gg 1$.

On Fig.2b we demonstrate the small differences $K(t) - K_0(t)$ and $S(t) - S_0(t)$ that are the corrections to the Schmidt number and the entanglement entropy compared to the values defined by Eqs. (18) and (19). Asymptotically for $\tau \gg 1$ and $\eta \ll 1$ the considered parameters are defined by:

$$K_\infty \approx 1 + \eta^2 \tag{20}$$

$$S_\infty \approx \frac{\eta^2}{\ln 2}\left(\ln\left(\frac{1}{\eta}\right) + 0.5(1 + \ln 2)\right) \tag{21}$$

We see that the residual asymptotical entanglement given by (20) and (21) is completely due to the energy spread of the initial wave packet.

The asymptotical properties of the atom-photon system can be clearly described by the momentum representation that is to be discussed in the next section.

### 4. Spontaneous radiation of an atom in momentum representation

According to the Weisskopf- Wigner theory the natural spectral line width of a two-level atom is defined by [6,9]:

$$a(\omega_{ph}) \sim \frac{1}{\omega_{ph} - \omega + i\gamma/2}, \tag{22}$$

there $\omega$ and $\gamma$ are the frequency and the width of the atomic transition, $\omega_{ph}$ - is the photon frequency and $a(\omega_{ph})$ is the probability amplitude.

Let us take into account momentum spread in initial wave packet and recoil energy of atom. As a result, we get:

$$\Psi(\omega_{ph}, p_a) = C \frac{\exp\left(-\dfrac{p_a^2 a_0^2}{2\hbar^2}\right)}{\omega_{ph}\left(1 - \dfrac{p_a}{Mc}\right) + \dfrac{\gamma}{2\xi_0} - \omega + i\gamma/2} \tag{23}$$

In the numerator stands the amplitude of the Gauss packet in impulse representation (Fourier transform of Eq. (10)). The multiplier $\left(1 - \frac{p_a}{Mc}\right)$ in the denominator describes the Doppler frequency shift, $\frac{\gamma}{2\xi_0}$ is the photon frequency decrease due to the atom recoil, $C$ accounts for normalization.

Eq. (23), as well as (9), defines the atom state for $t \gg 1/\gamma$, when the weight of the initial excited state is close to zero. We also suppose that during radiation time $1/\gamma$ the Gaussian atomic packet changes little (that corresponds to the condition $\eta \ll 1$).

Let us introduce dimensionless variables: the dimensionless photon frequency $\nu_{ph}$ (or equally its dimensionless momentum) and the dimensional atom momentum $\pi_a$.

$$\nu_{ph} = \frac{\omega_{ph} - \omega}{\gamma} \tag{24}$$

$$\pi_a = \frac{p_a a_0}{\hbar} \tag{25}$$

The Eq. (23) in dimensionless coordinates takes the form:

$$\Psi(\nu_{ph}, \pi_a) = C \frac{\exp\left(-\frac{\pi_a^2}{2}\right)}{\nu_{ph} + \frac{1}{2\xi_0} - \eta \pi_a + i/2} \tag{26}$$

In the considered approximation $\eta \ll 1$, $\xi_0 \gg 1$. Therefore the denominator of (26) is close to the denominator of the initial equation (22) of Weisskopf-Wigner. In that case the state (26) is close to the product of the Lorenzian photon multiplier (22) and the Gaussian atomic multiplier. In other words, the entanglement between $\nu_{ph}$ and $\pi_a$ is small. The deviation of the Schmidt number from one and the entanglement entropy from zero is defined by (20) and (21) that is approved by numerical calculations.

On Fig. 3a, 3b, 3c we show the structure of the first two Schmidt modes. Fig. 3a and 3b demonstrate probability densities and Fig. 3c – the real and the imaginary amplitude parts of photonic modes.

### 5. Spontaneous parametric down conversion of light

Spontaneous parametric down conversion (SPDC) of light is a non-linear optical effect that implies that the initial photon (pump laser generated) is transformed into two photons – signal and idler photons, the frequencies sum of which is equal to the pump frequency in stationary case.

There are two types of SPDC – with the first type phase matching and with the second type phase matching. In the first case the generated photons are polarized equally, while in the second case the photons are polarized perpendicularly to one another. Let us consider the latter case of the type-II SPDC. Then the generated photons can be associated with the ordinary and the extraordinary rays in a non-linear crystal. Let us consider a so called frequency-degenerated regime, when each of the generated photons have approximately equal frequencies. Due to the fact that the properties of the rays are different the photons become distinguishable. The effect is considerable for short pump impulses (10-100 fs), i.e. for broadband pump $10^{13} - 10^{14}$ Hz.

Let us consider the collinear mode that stands for the case when the directions of all the three photons are the same (the pump photon and the two generated photons). In [11-13] it was shown that for the Gaussian pump the probability amplitude $\psi(\omega_o, \omega_e)$ of generating a pair of photons with frequencies $\omega_o$ and $\omega_e$ for the ordinary and the extraordinary rays respectively can be expressed as:

$$\psi(\omega_o, \omega_e) = C \cdot \alpha(\omega_o, \omega_e) \cdot \Phi(\omega_o, \omega_e) \tag{27}$$

where $C$ is the normalization constant.

$$\alpha(\omega_o, \omega_e) = \exp\left(-\frac{(\omega_o + \omega_e - 2\bar{\omega})^2}{\sigma^2}\right) \tag{28}$$

$$\Phi(\omega_o, \omega_e) = \frac{\sin\left[\left(k_o(\omega_o) + k_e(\omega_e) - k_p(\omega_o + \omega_e)\right)\frac{L}{2}\right]}{\left[\left(k_o(\omega_o) + k_e(\omega_e) - k_p(\omega_o + \omega_e)\right)\frac{L}{2}\right]} \tag{29}$$

The first multiplier $\alpha(\omega_o, \omega_e)$ is defined by the pump spectrum. The parameter $2\bar{\omega}$ stands for the mean pump frequency value and $\sigma$ - is the pump frequency spread. Note that the pump frequency distribution has a variance $\sigma^2/4$.

The second multiplier $\Phi(\omega_o, \omega_e)$ is defined by the non-linear crystal properties. $L$ is the crystal length, $k_o(\omega_o)$, $k_e(\omega_e)$ and $k_p(\omega_o + \omega_e)$ are dispersion dependencies for the ordinary, extraordinary and the pump waves respectively.

If one considers the dispersion dependencies around the frequency $\bar{\omega}$ for the ordinary and the extraordinary waves and around the frequency $2\bar{\omega}$ for the pump wave then the multiplier $\Phi(\omega_o, \omega_e)$ can be approximately expressed as [11,12]:

$$\Phi(\omega_o, \omega_e) = \frac{\sin\left[\left((\omega_o - \bar{\omega})(k'_o - k'_p) + (\omega_e - \bar{\omega})(k'_e - k'_p)\right)\frac{L}{2}\right]}{\left[\left((\omega_o - \bar{\omega})(k'_o - k'_p) + (\omega_e - \bar{\omega})(k'_e - k'_p)\right)\frac{L}{2}\right]} \tag{30}$$

There the derivatives of the wave numbers on frequency define the inversed group wave velocities:

$$k'_o = \frac{\partial k_o(\omega)}{\partial \omega}\Big|_{\omega=\bar{\omega}} \quad k'_e = \frac{\partial k_e(\omega)}{\partial \omega}\Big|_{\omega=\bar{\omega}} \quad k'_p = \frac{\partial k_p(\omega)}{\partial \omega}\Big|_{\omega=2\bar{\omega}} \tag{31}$$

In calculations it is convenient to describe the spectrum width in inversed picoseconds $\sigma \sim (10-100) ps^{-1}$. Also, the product of the differences of the inversed group velocities and the crystal length is also convenient to be expressed in picoseconds. The typical experimental values that are used in the calculations are as follows [3]: $(k'_p - k'_e)L = 0.266\ ps$; $(k'_p - k'_o)L = 0.076\ ps$ (for the length of crystal $L = 1\ mm$).

The probability amplitude can be expressed in compact form if one introduces dimensionless variables and parameters:

$$p = \frac{\omega_o - \bar{\omega}}{\sigma}; \qquad q = \frac{\omega_e - \bar{\omega}}{\sigma} \tag{32}$$

$$X_o = (k'_p - k'_o)L\sigma; \qquad X_e = (k'_p - k'_e)L\sigma \tag{33}$$

Then the probability amplitude of SPDC can be expressed as:

$$\psi(\omega_o, \omega_e) = C \cdot \exp\left(-(p+q)^2\right) \frac{\sin\left[0.5(X_o p + X_e q)\right]}{\left[0.5(X_o p + X_e q)\right]}$$

$$\psi(\omega_o, \omega_e) = C \cdot \exp\left(-(p+q)^2\right) \frac{\sin\left[0.5(X_o p + X_e q)\right]}{\left[0.5(X_o p + X_e q)\right]} \tag{34}$$

**6. The coherence of the photons that make a bi-photon**

Let the radiation field of bi-photons (e.g. after beam splitter passing) be located in two different spatial modes (the signal mode and the idler mode).

Let the ordinary ray be horizontally polarized and the extraordinary ray vertically-polarized. Then the vector of the bi-photon state generated during the type-II SPDC includes the frequency and the polarization degrees of freedom and can be expressed in the form:

$$|\Psi\rangle = \frac{1}{\sqrt{2}}\left[\psi(\omega_o^s, \omega_e^i)|HV\rangle + \psi(\omega_e^s, \omega_o^i)|VH\rangle\right] \tag{35}$$

There the state vector consists of two summands: the probability amplitude of the ordinary ray to be in a signal mode and the extraordinary in the idler mode and vice versa.

Let $|H\rangle = \begin{pmatrix} 1 \\ 0 \end{pmatrix}$, $|V\rangle = \begin{pmatrix} 0 \\ 1 \end{pmatrix}$, $|e_1\rangle = |HV\rangle = \begin{pmatrix} 0 \\ 1 \\ 0 \\ 0 \end{pmatrix}$, $|e_2\rangle = |VH\rangle = \begin{pmatrix} 0 \\ 0 \\ 1 \\ 0 \end{pmatrix}$. Then in dimensionless discrete representation the bi-photon state (35) can be expressed as:

$$|\Psi\rangle = \frac{1}{\sqrt{2}}\left[\psi_{pq}|e_1\rangle + \psi_{qp}|e_2\rangle\right] \tag{36}$$

The frequency parts of the summands of amplitudes $\psi_{pq}$ and $\psi_{qp}$ are symmetrically conjugated matrices obtained for the discretization of the Eq. (34).

The state vector (36) is a ket-vector. The conjugated to it vector (a bra-vector) is:

$$\langle\Psi| = \frac{1}{\sqrt{2}}\left[\psi_{pq}^*\langle e_1| + \psi_{qp}^*\langle e_2|\right] \tag{37}$$

When one registers the polarization of photons and does not measure the frequencies then one gets a polarization density matrix. It is obtained by summing the complete density matrix for the frequency degrees of freedom, according to the standard rules of quantum mechanics [14]:

$$\hat{\rho} = Tr_{pq}\left(|\Psi\rangle\langle\Psi|\right) \tag{38}$$

The results of our calculations can be expressed in the following compact form:

$$\hat{\rho} = \frac{1}{2}\begin{pmatrix} 0 & 0 & 0 & 0 \\ 0 & 1 & F & 0 \\ 0 & F^* & 1 & 0 \\ 0 & 0 & 0 & 0 \end{pmatrix} \tag{39}$$

There the parameter $F$, that characterizes the degree of the state coherence, is given by:

$$F = \sum_{pq}\psi_{pq}\psi_{qp}^* = Tr(\psi\psi^*) \tag{40}$$

Note that in the aforementioned equation $\psi^*$ one has a complex rather than the Hermitian matrix conjugation. Also note that $Tr(\psi\psi^+) = 1$ due to the normalization condition.

In our matter the parameter $F$ is a real number ($0 \leq F \leq 1$). For $F = 1$ one gets a pure state $\frac{1}{\sqrt{2}}[|HV\rangle + |VH\rangle] = \begin{pmatrix} 0 \\ 1/\sqrt{2} \\ 1/\sqrt{2} \\ 0 \end{pmatrix}$. Contrary, for $F = 0$, one gets a completely incoherent mixture of states $|HV\rangle$ and $|VH\rangle$ with equal weights. If $0 < F < 1$ then one gets a partially coherent polarization state, that consists of a mixture of the state $\frac{1}{\sqrt{2}}[|HV\rangle + |VH\rangle] = \begin{pmatrix} 0 \\ 1/\sqrt{2} \\ 1/\sqrt{2} \\ 0 \end{pmatrix}$ with the weight $\frac{1+F}{2}$ and the state $\frac{1}{\sqrt{2}}[|HV\rangle - |VH\rangle] = \begin{pmatrix} 0 \\ 1/\sqrt{2} \\ -1/\sqrt{2} \\ 0 \end{pmatrix}$ with the weight $\frac{1-F}{2}$.

The above-described theoretical analysis can also be performed experimentally by the methods of bi-photon fields quantum tomography. [15,16].

On Fig. 4 the dependence of the coherence degree $F$ on the crystal length is shown (for the pump bandwidth $\sigma = 10\,ps^{-1}$). Note that according to (33) the calculation results depend on the product of the crystal length and the pump frequency spread $\sigma$. On account of this, the Fig. 4 can be easily recalculated for other values of $\sigma$.

Fig. 5 illustrates the closeness of the ordinary and the extraordinary modes to one another for the higher values of coherence $F$ (or equally for the smaller crystal lengths or the pump bandwidth). The corresponding odd modes are aloes and the even modes differ in sign. The figure corresponds for the following values of parameters:

$$L = 0.5\,mm, \sigma = 10\,ps^{-1}, F = 0.97, K = 5.6, S = 3.16.$$

Similarly, Fig.6 shows the difference in modes for the ordinary and the extraordinary rays for the lower values of coherence $F$. The figure corresponds for the values of parameters:

$$L = 4\,mm, \sigma = 10\,ps^{-1}, F = 0.37, K = 2.2, S = 1.8$$

### 7. Conclusions

Let us briefly summarize the main results of our present work.

1. The algorithm of Schmidt modes extraction is applied to the systems with continuous physical variables. The algorithm is based on a singular value matrix decomposition. The simplicity, the high accuracy and the high reliability are demonstrated. The comparison with other methods is performed.

2. It is shown that a parameter based on fundamental constants defines the evolution of the entangled states of the atom-photon system with spontaneous radiation. The parameter consists of the product of the fine structure constant and the atom-electron mass ratio.

3. The analysis of dynamics of entanglement in the system atom-photon is performed. It is shown that in the course of time the entropy of entanglement between the photonic and the atomic degrees of freedom increases for the times of the order of magnitude of the excited state life-time. Then the entropy falls and asymptotically approaches to the small residual entropy level that is defined by the energy spread of the initial wave packet. The analytical structure of the photonic modes is obtained.

4. The effect of the coherence loss between modes in the process of spontaneous parametric down conversion with second type phase matching. The polarization density matrix is described and the coherence parameter that is due to the difference between the ordinary and the extraordinary waves in a non-linear crystal is introduced. The dependence of the coherence degree between modes on the crystal length and the pump bandwidth is analyzed. Some examples of the Schmidt modes analysis for various coherence degrees are given.

Useful discussions with B.A. Grishanin and S.P. Kulik are gratefully acknowledged.

**References**

1. *Valiev K.A., Kokin A.A.* Quantum computers: reliance and reality. Second edition. Moscow- Izhevsk. R&C Dynamics. 2004. (in Russian). 320 c.
2. *Chan K.W., Law C.K., Eberly J.H.* Localized single- photon wave function in free space // Phys. Rev. Lett. 2002. V.88. 100402. 4p.
3. *Law C.K., Walmsley I.A., Eberly J.H.* Continuous frequency entanglement: effective finite Hilbert space and entropy control // Phys. Rev. Lett. 2000. V.84. P.5304-5307.
4. *Lamata L., Leon J.* Dealing with entanglement of continuous variables: Schmidt decomposition with discrete sets of orthogonal functions // LANL Report quant-ph/0410167. 2004. 6p.
5. *Fedorov M.V., Efremov M.A., Kazakov A.E. et al.* Spontaneous emission of a photon: wave packet structures and atom-photon entanglement // LANL Report quant-ph/0412107. 2004. 30p.
6. *Berestetskii V.B., Lifshitz E.M., Pitaevskii L.P.* Relativistic Quantum Theory 1. Pergamon Press. 1971.
7. *Balashov V.V., Dolinov V.K.* Quantum mechanics. Moscow State University Press. 1982 (in Russian).
8. *Chan K.W., Law C.K., Eberly J.H.* Quantum entanglement in photon- atom scattering // Phys. Rev. A. 2003. V.68, 022110, 9p.
9. *Scully M.O., Zubairy M.S.* Quantum Optics. Cambridge, University Press. 1997.
10. *Klyshko D.N.* Photons and Nonlinear Optics .Gordon & Breach. New York. 1988.
11. *Grice W.P., Walmsley I.A.* Spectral information and distinguishability in type- II down- conversion with a broadband pump // Phys. Rev. A. 1997. V.56. P.1627- 1634.
12. *Kim Y.H., Chekhova M.V., Kulik S.P. et al.* First- order interference of nonclassical light emitted spontaneously at different times // Phys. Rev. A. 2000. V.61, 051803, 4p.
13. *Keller T.E. and Rubin M.* Theory of Two-Photon Entanglement for Spontaneous Down-Conversion driven by a Narrow Pump Pulse // Phys. Rev. A. 1997. V.56. P. 1534-1541.
14. *Landau L.D., Lifschitz E.M.* Quantum mechanics (non- relativistic theory). 3$^{rd}$ ed. Pergamon Press. Oxford. 1991.
15. *Bogdanov Yu.I., Chekhova M.V., Kulik S.P. et al.* Statistical reconstruction of qutrits // Phys. Rev. A. 2004. V.70, 042303, 16p.
16. *Bogdanov Yu.I., Chekhova M.V., Kulik S.P. et al.* Qutrit State Engineering with Biphotons// Phys. Rev. Lett. 2004. V.93. 230503. 4p.

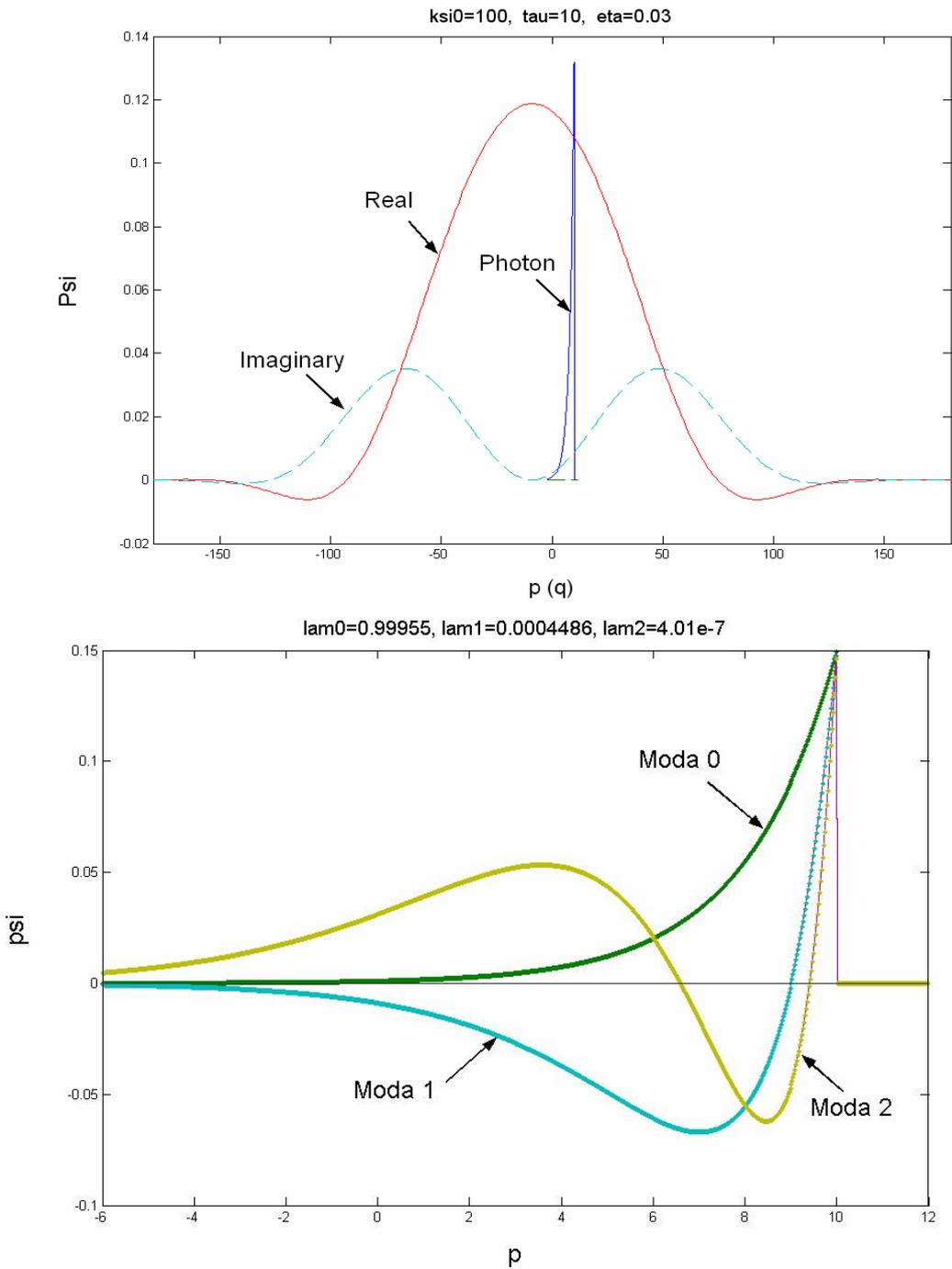

Fig. 1a: an example of probability amplitudes for the compound atom-photon system (the first Schmidt mode).
Fig 1b: the structure of the primary three wave functions of the radiation. There the greatest part stands for the first Schmidt mode.

The figure corresponds to the following parameter values: $\xi_0 = 100$, $\eta = 0.03$, $\tau = 10$.

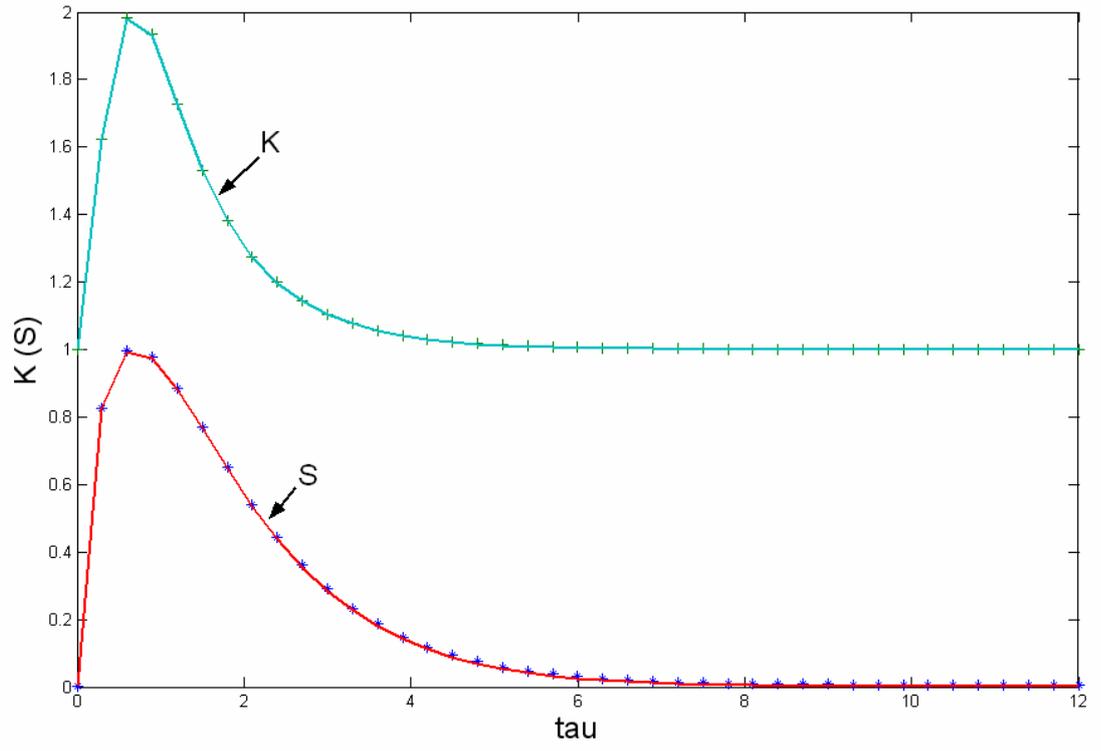

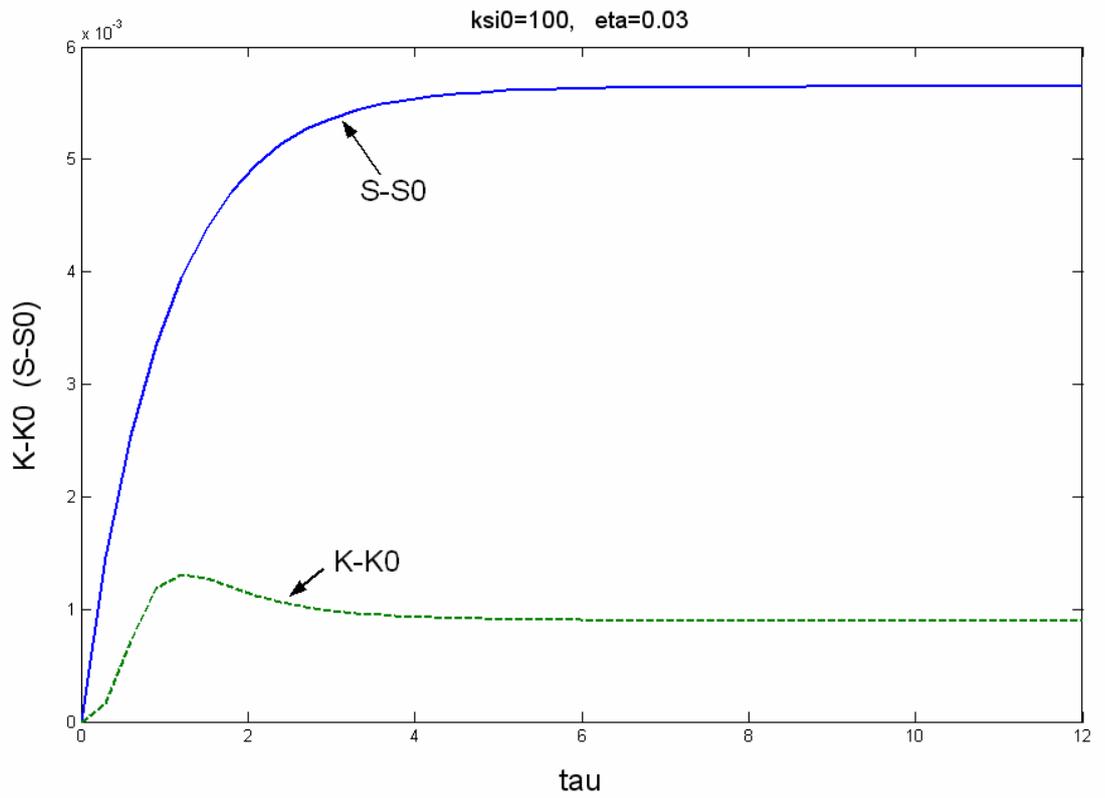

Fig. 2a: numerical calculations of the Schmidt number and the entanglement entropy.
Fig.2b: the small differences $K(t) - K_0(t)$ and $S(t) - S_0(t)$.

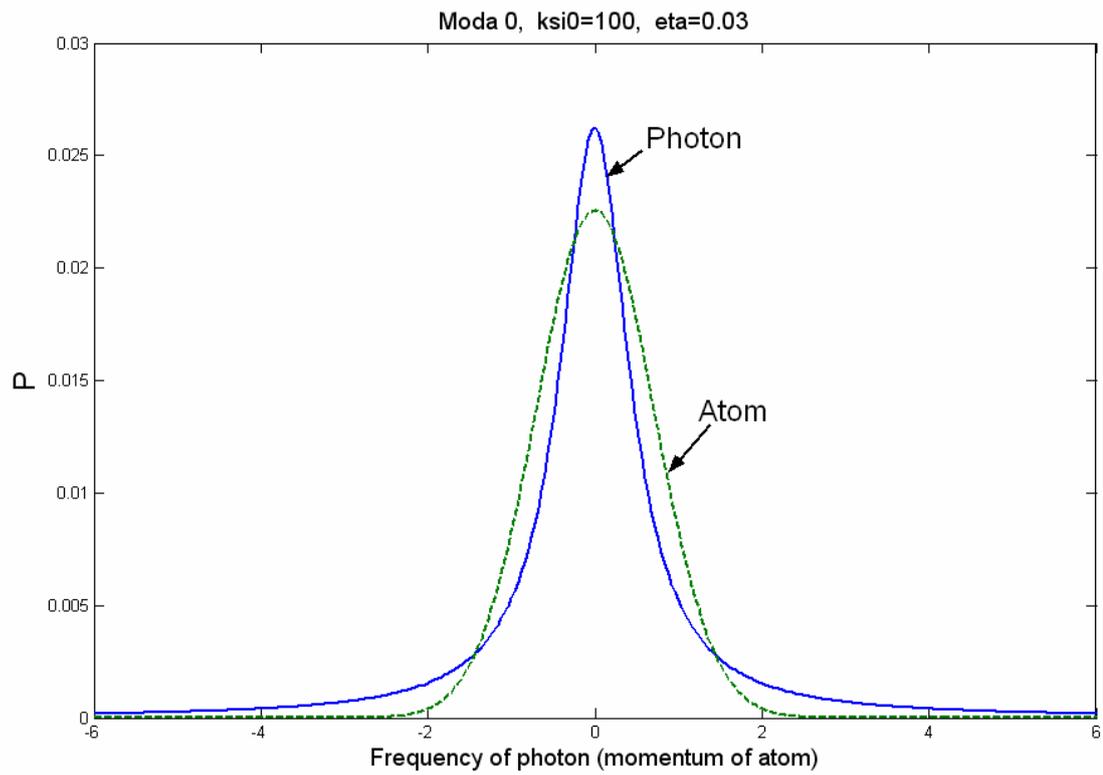

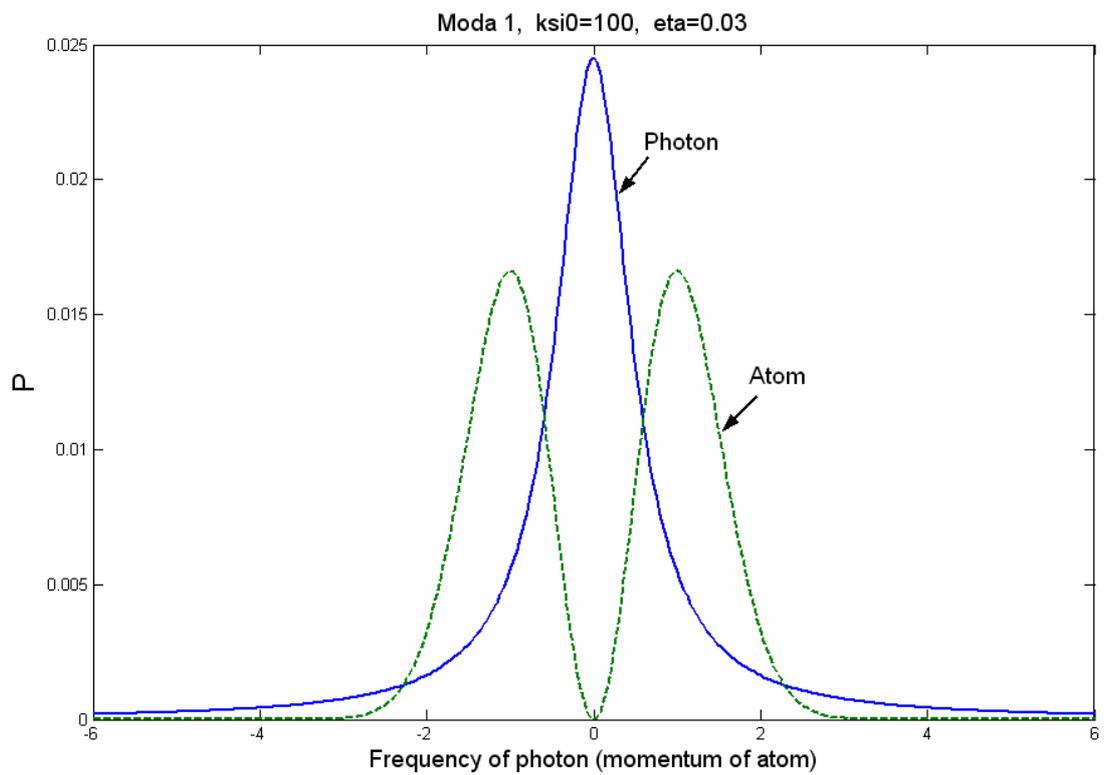

Fig. 3a, 3b: the structure of the first two Schmidt modes (probability densities)

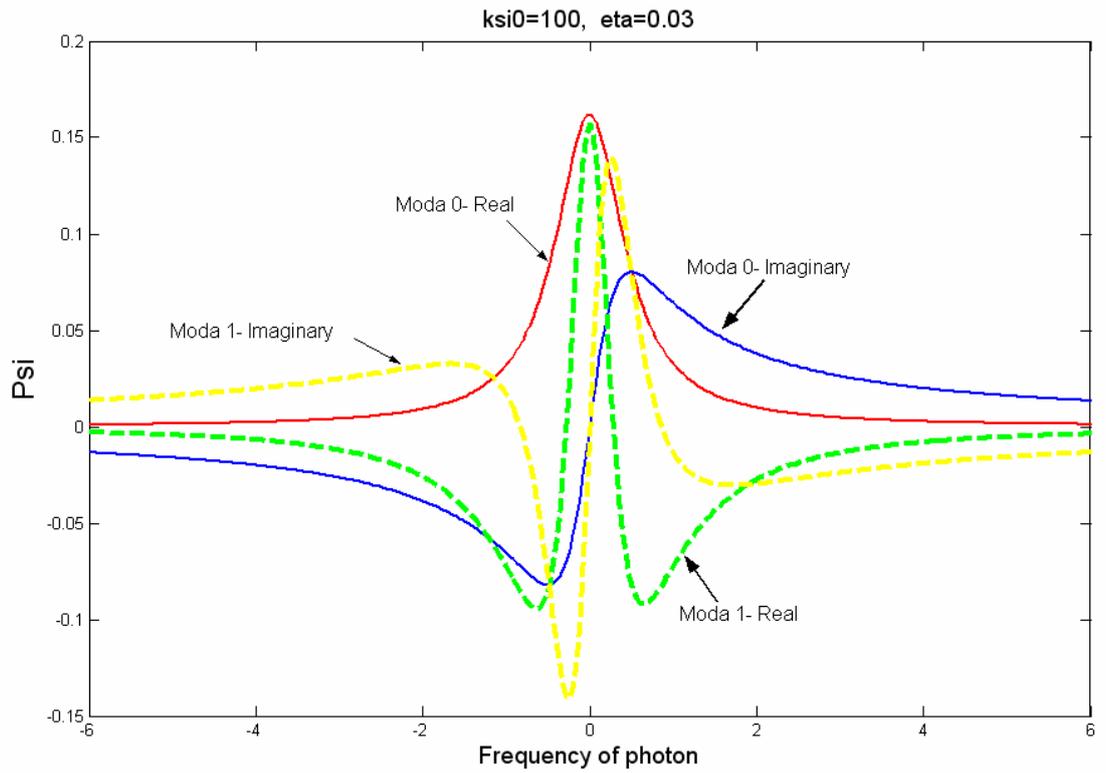

Fig. 3c: the structure of the first two Schmidt modes (the real and the imaginary amplitude parts of photonic modes)

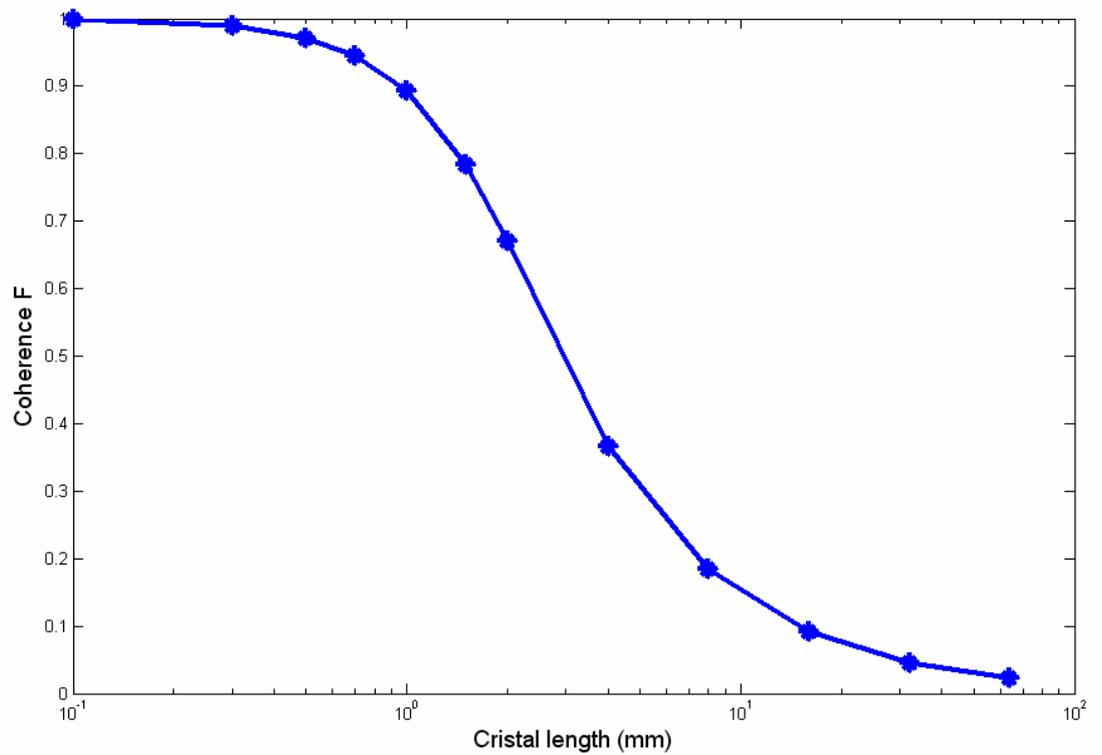

Fig. 4: the dependence of the coherence degree $F$ on the crystal length (for the pump bandwidth $\sigma = 10\, ps^{-1}$).

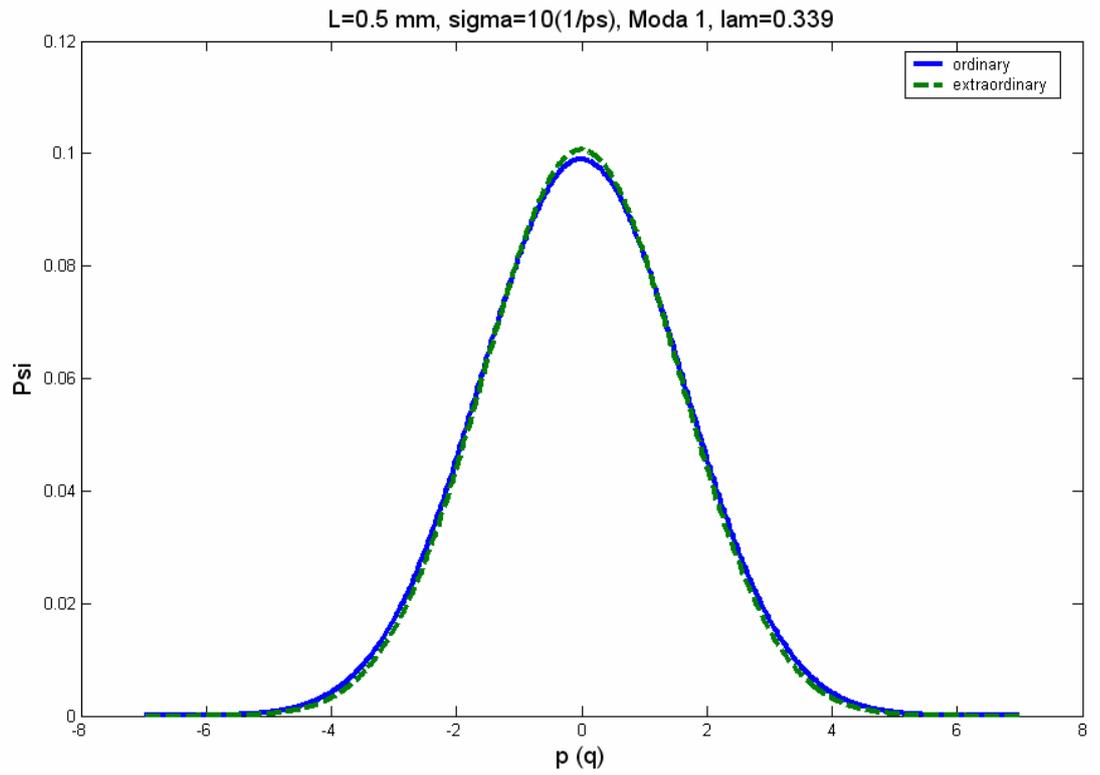

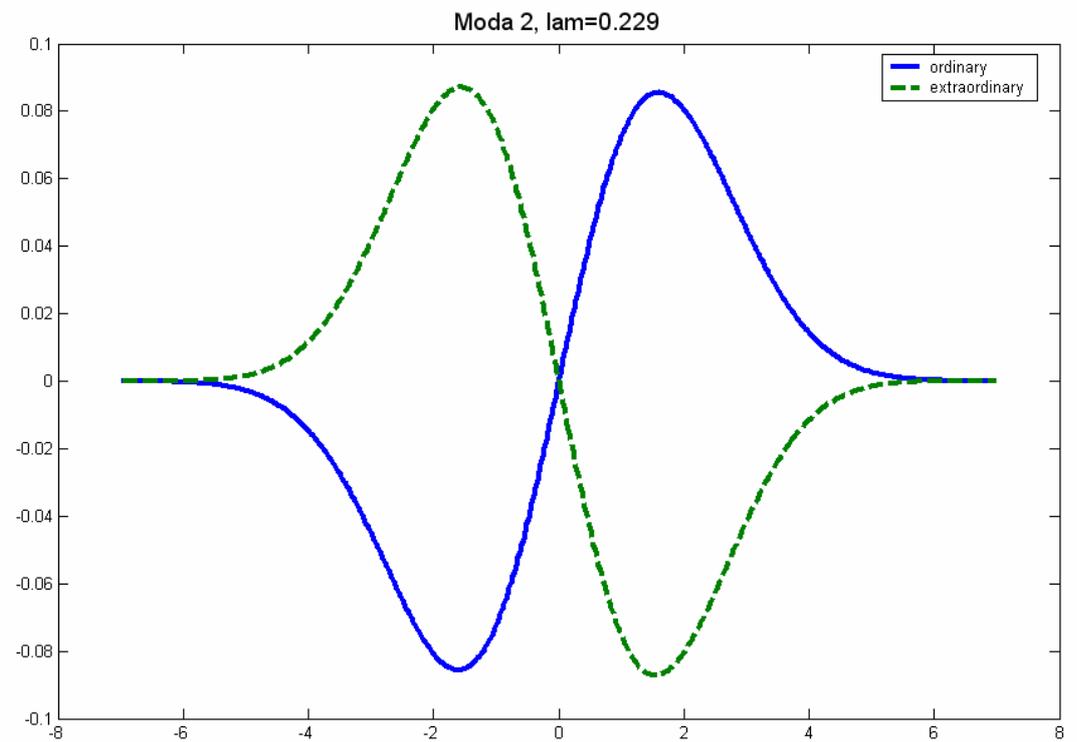

Fig. 5a, 5b: the closeness of the ordinary and the extraordinary modes to one another for the higher values of coherence $F = 0.97$ (or equally for the smaller crystal lengths or the pump bandwidth). Modes 1 and 2.

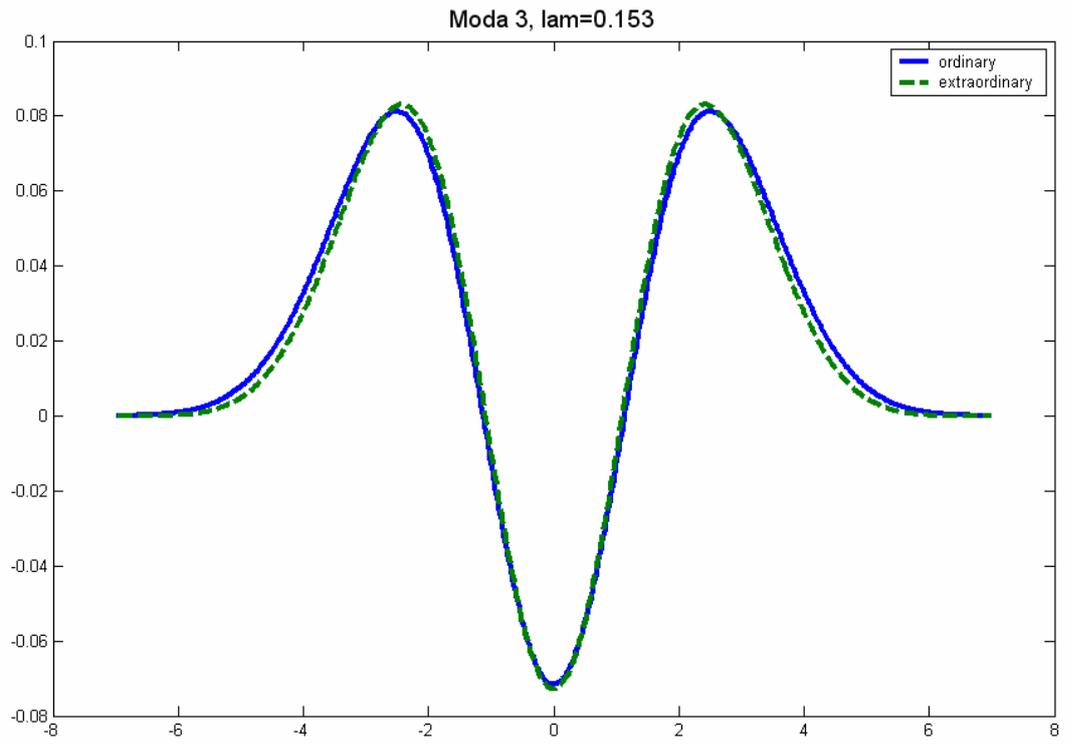

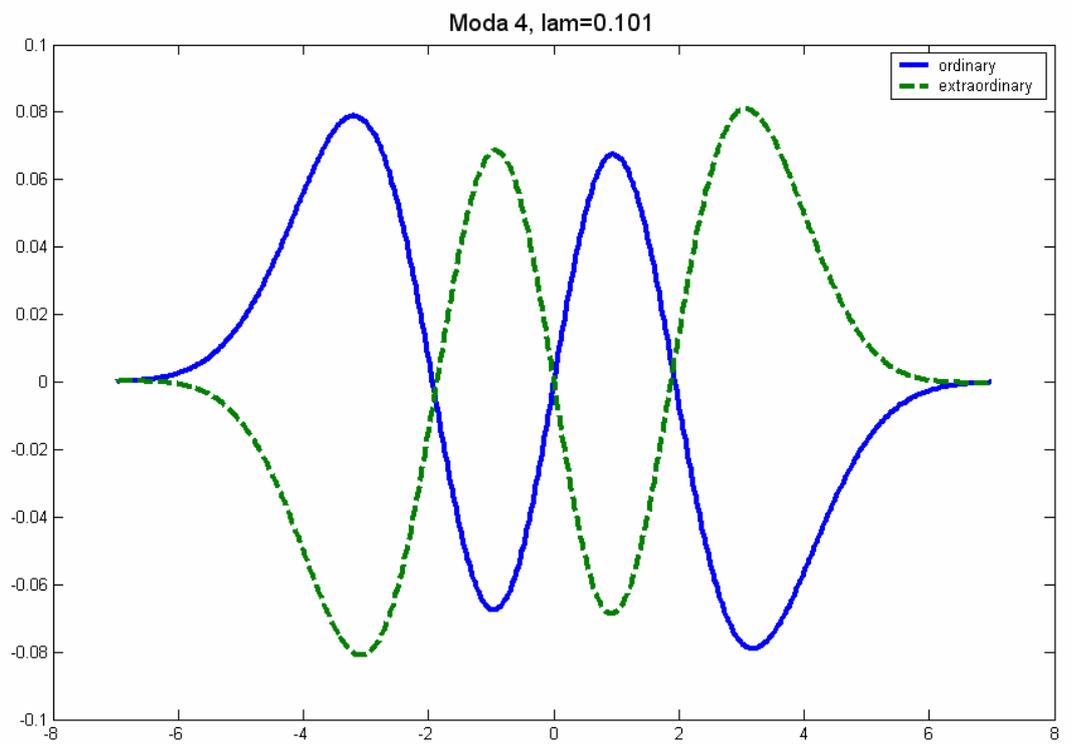

Fig. 5c, 5d: the closeness of the ordinary and the extraordinary modes to one another for the higher values of coherence $F = 0.97$ (or equally for the smaller crystal lengths or the pump bandwidth). Modes 3 and 4.

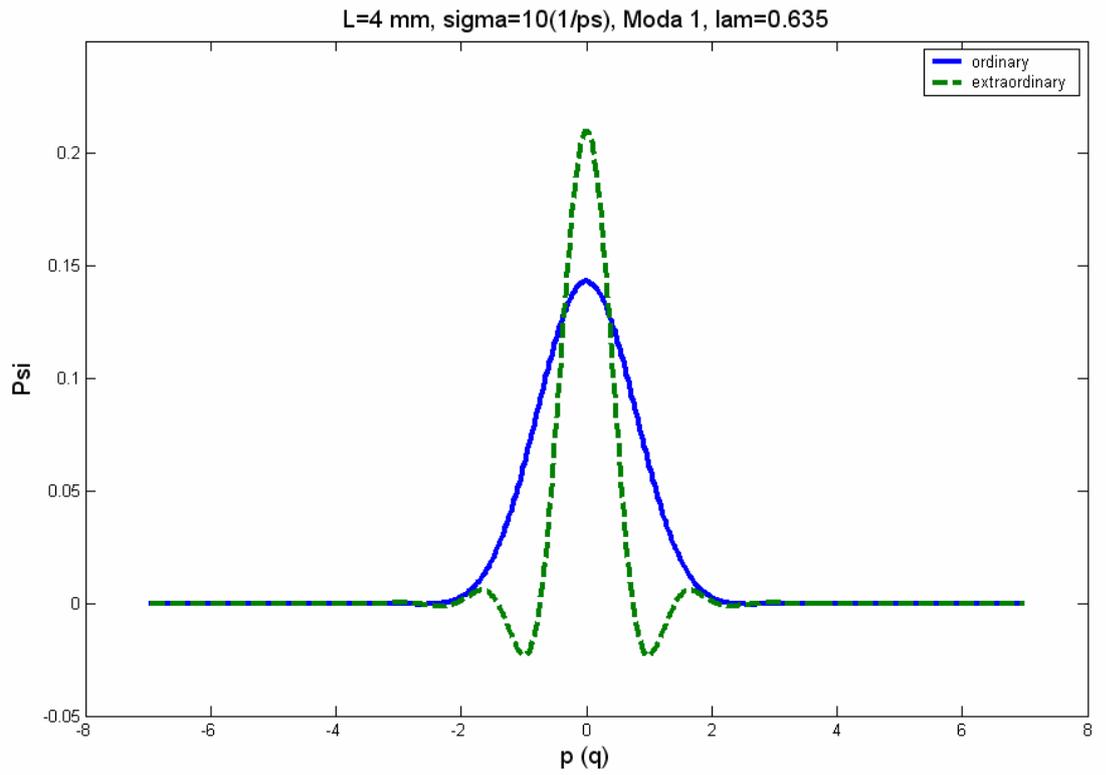

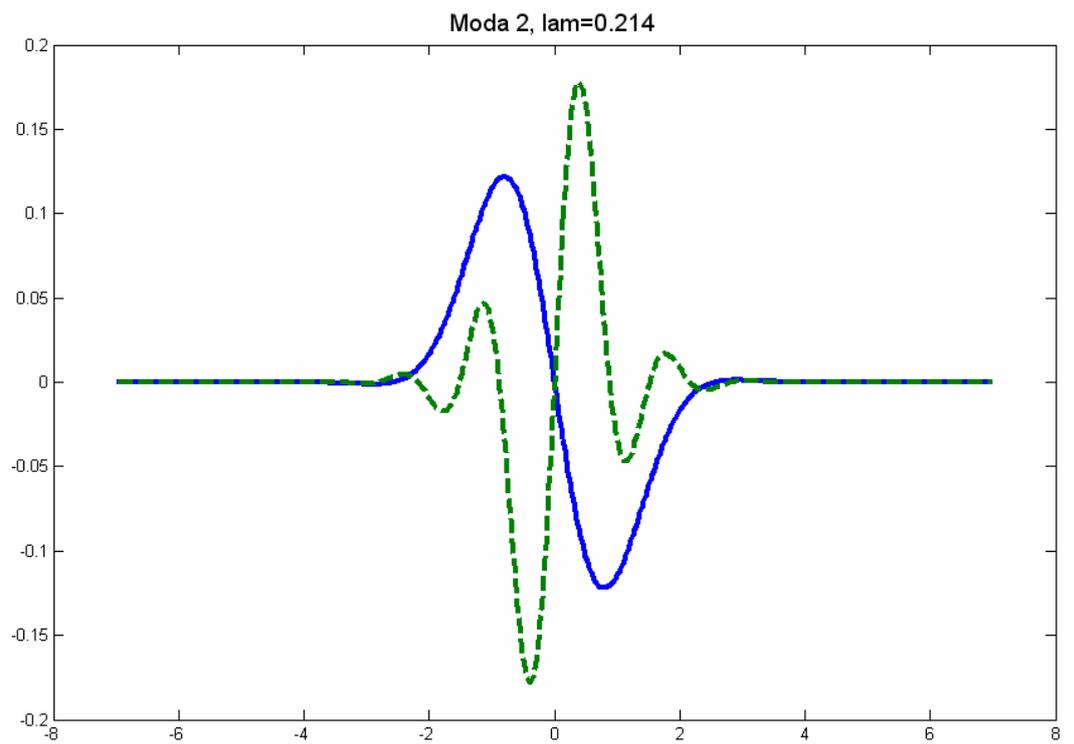

Fig.6a,6b: the difference in modes for the ordinary and the extraordinary rays for the lower values of coherence $F = 0.37$. Modes 1 and 2.

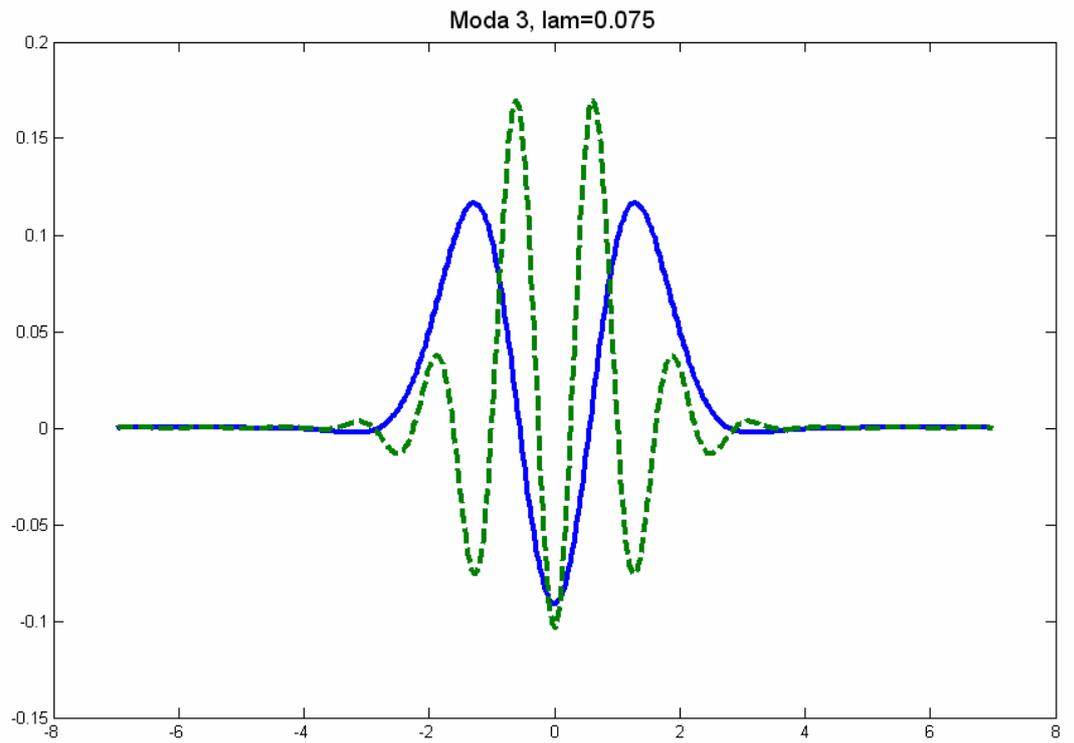

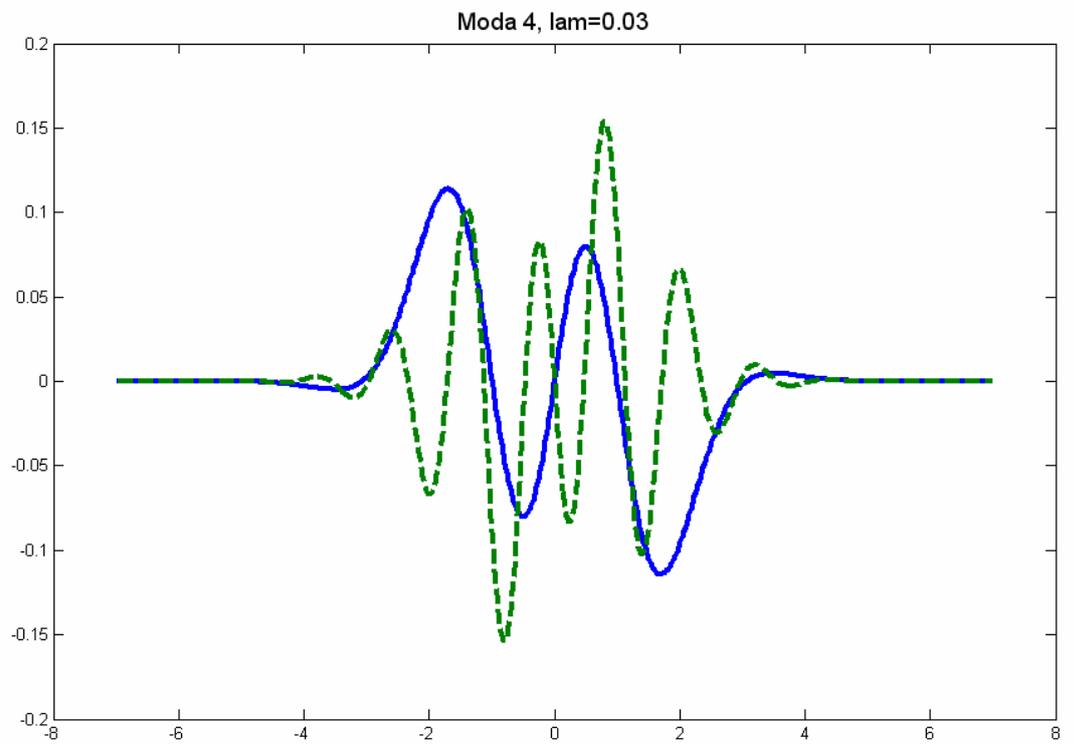

Fig.6c,6d: the difference in modes for the ordinary and the extraordinary rays for the lower values of coherence $F = 0.37$. Modes 3 and 4.